\documentclass[10pt,prd,preprintnumbers,floatfix,aps,nofootinbib,twocolumn,superscriptaddress,noshowpacs]{revtex4-2}
\usepackage[utf8]{inputenc}
\usepackage[T1]{fontenc}
\usepackage{graphicx}
\usepackage{amsmath}
\usepackage{amssymb}
\usepackage[dvipsnames]{xcolor}
\usepackage{silence}
\usepackage{newtxtext}
\usepackage{newtxmath}
\usepackage[cal=cm]{mathalfa}

\usepackage[colorlinks]{hyperref}
\AtBeginDocument{\hypersetup{
        linkcolor=Blue,
        allcolors=Blue
}}

\usepackage{etoolbox}
\usepackage{array}
\AtBeginEnvironment{table}{\setlength{\abovecaptionskip}{2pt}}
\AtBeginEnvironment{table*}{\setlength{\abovecaptionskip}{2pt}}

\usepackage{overpic}
\usepackage{refcount}
\newcommand{\figeqlink}[4]{%
  \put(#1,#2){\small\hyperref[#3]{\strut\phantom{#4}}}%
  \ifnum\getrefnumber{#3}=#4\relax\else
    \PackageWarningNoLine{figeqlink}{Figure prints (#4) for #3, but the
      paper now numbers it (\getrefnumber{#3}). Update the figure's .tex}%
  \fi}
\newcommand{\figsteptarget}[3]{\put(#1,#2){\raisebox{\baselineskip}{\hypertarget{#3}{}}}}
\newcommand{\figsteplink}[5]{\put(#1,#2){\hyperlink{#5}{\makebox(#3,#4){}}}}

\begin{document}
\preprint{IFT-UAM/CSIC-26-173}

\title{A Genetic Algorithm Approach to Reconstructing Scalar-Tensor Lagrangians} 
%%%%%

\author{Mikel Artola}
\email{mikel.artola@ehu.eus}
\affiliation{Department of Physics, University of the Basque Country EHU, 48940 Leioa, Spain}
\affiliation{EHU Quantum Center, University of the Basque Country EHU, 48940 Leioa, Spain}

\author{George Alestas}
\email{g.alestas@csic.es}
\affiliation{Instituto de F\'isica Te\'orica (IFT) UAM-CSIC, C/ Nicol\'as Cabrera 13-15, Campus de Cantoblanco UAM, 28049 Madrid, Spain}

\author{Dimitrios Efstratiou}
\email{d.efstratiou@uoi.gr}
\author{Evangelos Achilleas Paraskevas}
\email{e.paraskevas@uoi.gr}
\affiliation{Department of Physics, University of Ioannina, GR-45110, Ioannina, Greece}

\begin{abstract}
Recent DESI baryon acoustic oscillation (BAO) data, combined with cosmic microwave background (CMB) and Type Ia supernovae observations, have been interpreted as evidence for dynamical dark energy whose equation of state crosses the phantom divide $w=-1$. Such a crossing cannot be realized smoothly by a minimally coupled scalar field $\varphi$, but can arise naturally in scalar-tensor theories. We present a data-driven framework that reconstructs the scalar-tensor Lagrangian directly from observations, using a genetic algorithm to evolve the symbolic expressions for the nonminimal coupling $F(\varphi)$ and the scalar potential $U(\varphi)$. Each candidate is evaluated through a forward model that self-consistently determines the expansion history, the effective gravitational coupling, and the growth of structure. Candidates are ranked by a joint likelihood of DESI DR2 BAO, Pantheon+ supernovae calibrated with SH0ES Cepheids, compressed CMB distance priors, and redshift-space distortion measurements of $f\sigma_8$. We restrict to the unscreened branch of the theory and impose Solar System constraints on the post-Newtonian parameter $\gamma$ and on the time variation of the gravitational coupling, together with a Big Bang nucleosynthesis bound on its evolution. Across 64 independent runs, the best reconstruction improves the fit relative to $\Lambda$CDM by $\Delta\chi^2=-16.6$, compared with $\Delta\chi^2=-10.6$ for the CPL dark-energy parametrization. The preferred coupling is well described by $F(\varphi)=c_1\varphi^2+e^{\varphi}(1-c_2\varphi)$. The reconstructed theory shows phantom behavior at $z\gtrsim0.5$, but it remains close to $w=-1$ at low redshift.
\end{abstract}

\maketitle

\section{\label{sec:introduction}Introduction}

The \(\Lambda\)CDM model remains the standard framework for describing the cosmic expansion history and the formation of large-scale structure, with remarkable success across a broad range of observations~\cite{Planck:2018vyg, eBOSS:2020yzd, Brout:2022vxf, DESI:2025zgx}. Nevertheless, persistent discrepancies have raised questions about its completeness~\cite{2008arXiv0811_4684P, Perivolaropoulos:2021jda, Abdalla:2022yfr, Efstathiou:2024dvn, Peebles:2024txt, Peebles:2022akh, Bull:2015stt, Bullock:2017xww, DiValentino:2017gzb, CosmoVerseNetwork:2025alb}. Among the most widely discussed are the Hubble-constant tension, namely the mismatch between early- and late-Universe determinations of \(H_0\)~\cite{Planck:2018vyg, Riess:2021jrx, DiValentino:2021izs, Hu:2023jqc, Freedman:2017yms, DiValentino:2020zio, Verde:2019ivm, Shah:2021onj, Schoneberg:2021qvd, Verde:2023lmm, Perivolaropoulos:2024yxv, Pantos:2026cxv}, and the milder but still actively debated tension in the amplitude of matter clustering, commonly quantified through \(S_8\)~\cite{Heymans:2020gsg, DES:2021wwk, Wright:2025xka, DiValentino:2020vvd, Pantos:2026koc}. These anomalies have motivated extensive exploration of extensions of \(\Lambda\)CDM and, more generally, of new dynamical degrees of freedom beyond the standard cosmological model~\cite{Mortonson:2013zfa}.

A particularly important development in this context is the recent indication of dynamical dark-energy behavior from the Dark Energy Spectroscopic Instrument (DESI) baryon acoustic oscillation (BAO) observations~\cite{DESI:2024mwx, DESI:2025zgx, DESI:2025fii, DESI:2024aqx, DESI:2024hhd, DESI:2025wyn, RoyChoudhury:2024wri, RoyChoudhury:2025iis}. In analyses performed within General Relativity (GR), the DESI DR2 data, combined with cosmic microwave background (CMB)~\cite{Planck:2018vyg} and Type Ia supernovae (SN) observations~\cite{Scolnic:2021amr, Rubin:2023jdq, DES:2024jxu}, favor an evolving dark-energy equation of state \(w(z)\), with indications of a possible crossing of the phantom divide \(w = -1\)~\cite{DESI:2025zgx, DESI:2025fii, Shajib:2025tpd, Specogna:2025guo, RoyChoudhury:2025dhe, Lu:2025gki, Li:2024qso, Li:2025owk, Cai:2025mas, Alestas:2025syk, Wang:2025xvi, Scherer:2025esj, Sabogal:2025jbo, Silva:2025twg}. The statistical significance of this preference, however, remains debated, since it depends on the supernova sample and on the dark-energy parametrization adopted, and Bayesian model comparison does not yield decisive evidence against \(\Lambda\)CDM~\cite{Cortes:2024lgw, Nesseris:2025lke, Wolf:2025jlc, Wang:2025bkk, Ong:2025utx, Ong:2026tta, LozanoTorres:2026xxx, Zapata:2026evt}. At the same time, such reconstructions tend to prefer a lower value of \(H_0\), thereby worsening the disagreement with local measurements~\cite{Pang:2025lvh,Mirpoorian:2025rfp}. These developments motivate the study of modified-gravity descriptions of late-time cosmic acceleration~\cite{Ye:2024ywg, Pan:2025psn, Wolf:2025jed, Tsujikawa:2025wca, Efstratiou:2025iqi, Argudo-Panes:2026qkw}.

Minimally coupled quintessence remains the simplest physically motivated model of dynamical dark energy, and has been confronted with DESI data in several forms, from single- and multi-field exponential potentials~\cite{Akrami:2025zlb, Alestas:2025syk} to quintessential inflation~\cite{Jing:2026ymp}. Since \(w \geq -1\) in these models, they cannot reproduce the phantom crossing suggested by CPL. Nevertheless, exponential quintessence was found to fit the data significantly better than \(\Lambda\)CDM, suggesting that the crossing may partly reflect the limitations of the Chevallier--Polarski--Linder (CPL) dark-energy parametrization ~\cite{Chevallier:2000qy, Linder:2002et} rather than a feature required by the data~\cite{Nesseris:2025lke}. More generally, a smooth phantom crossing is difficult to realize with a single minimally coupled scalar field~\cite{Wolf:2024eph, Park:2025fbl, Gialamas:2025pwv}. For a scalar field with Lagrangian \(\mathcal{L} = \pm \dot{\varphi}^2/2 - U(\varphi)\), the sign of the kinetic term is fixed, and a continuous transition across \(w = -1\) would require it to vanish and change sign dynamically, which cannot occur without pathologies~\cite{Vikman:2004dc, Carroll:2003st, Cline:2003gs, Perivolaropoulos:2005yv, Kunz:2006wc, Cai:2009zp}. This obstruction has motivated multi-component scenarios~\cite{Feng:2004ad, Guo:2004fq, Hu:2004kh, Chimento:2008ws, Saitou:2012xw, Chaves:2008gd, Cai:2025mas}, dark-sector interactions~\cite{Huey:2004qv, Das:2005yj, Khoury:2025txd, Zhang:2025dwu}, dynamical vacuum models~\cite{Sola:2005et, Sola:2016jky, Sola:2017znb, SolaPeracaula:2021yid}, or modifications of gravity itself~\cite{Boisseau:2000pr, Nesseris:2006er, Gannouji:2006jm}.

Scalar-tensor theories~\cite{Brans:1961sx, Singh:1987is, Faraoni:2004pi, Quiros:2019ktw, Horndeski1974, Langlois:2018dxi, Nojiri:2003ft, Papantonopoulos:2007zz} are especially relevant in this context because they can generate effective phantom-crossing behavior while maintaining a direct link between background evolution, gravitational coupling, and structure growth~\cite{Perivolaropoulos:2005yv, Gannouji:2006jm, Wolf:2024stt, Wolf:2025jed, Yao:2025wlx, SanchezLopez:2025uzw}. Previous reconstruction approaches have shown that scalar-tensor gravity can accommodate the DESI-motivated phantom-crossing behavior and alleviate the associated worsening of the \(H_0\) tension~\cite{Efstratiou:2025iqi}. In particular, the cosmological evolution can be used to reconstruct the nonminimal coupling \(F(\varphi)\) and the scalar potential \(U(\varphi)\)~\cite{Boisseau:2000pr,Esposito-Farese:2000pbo,Perivolaropoulos:2005yv,Capozziello:2007iu,Gannouji:2006jm}, suggesting that scalar-tensor gravity provides a viable and physically motivated framework for addressing the DESI DR2 phenomenology~\cite{Efstratiou:2025iqi,Ye:2024ywg,Pan:2025psn,Wolf:2025jed} and its connection to late-time cosmological tensions~\cite{Nesseris:2017vor,Kazantzidis:2019nuh}.

The present work takes the next step beyond such reconstruction programs. Rather than assuming specific ans\"atze for \(F(\varphi)\) and \(U(\varphi)\), or fitting preselected functional templates to reconstructed solutions, we aim to infer their functional structure directly from cosmological data. To achieve this, we combine a scalar-tensor forward model with a Genetic Algorithm (GA) that evolves symbolic candidate expressions for the coupling \(F(\varphi)\) and the potential \(U(\varphi)\), evaluates them against the observational likelihood, and progressively identifies the forms that best reproduce the data.

GAs constitute a class of stochastic optimization methods which, applied to symbolic regression, search for analytic functions that describe a given set of observational data. These algorithms, together with related symbolic-regression techniques, have been extensively used in cosmology~\cite{Bogdanos:2009ib, Nesseris:2010ep, Nesseris_2012, Arjona:2020kco, Euclid:2021frk, Arjona:2021hmg, Aizpuru:2021vhd, Orjuela-Quintana:2022nnq, Gomez-Vargas:2022bsm, Gangopadhyay:2023nli, Orjuela-Quintana:2023uqb, Bartlett:2022kyi, Medel-Esquivel:2023nov, Orjuela-Quintana:2024hha, Arjona:2024dsr, Arjona:2024cex, Peronaci:2026iur, DeLeo:2026luv}, as well as in several other areas of physics and beyond~\cite{Akrami:2009hp, Bao:2017thx, Udrescu:2019mnk, Luo:2019qbk, Acampora:2023pti, Tenachi_2023, Bartlett:2023gvh, rs12050866, AbdusSalam:2024obf}. The main advantage of this approach is that the data guide the functional form of the reconstructed quantities, rather than restricting the analysis to a small set of predefined ans\"atze. Unlike other non-parametric approaches (see e.g.~Refs.~\cite{Escamilla:2021uoj, Escamilla:2023shf}), GAs return closed-form analytic expressions.  In our approach, all cosmological observables are generated self-consistently from a single candidate pair \(\{F(\varphi),U(\varphi)\}\) through the background equations of scalar-tensor gravity. This enforces consistency among the expansion history, the effective gravitational coupling, and the growth of matter perturbations, while yielding directly a reconstructed Lagrangian rather than only an effective phenomenological description over a limited redshift range.

The main objective of this study is therefore to develop and implement a data-driven framework for reconstructing scalar-tensor Lagrangians from cosmological observations using GAs. More specifically, we seek to determine whether the combined DESI DR2~\cite{DESI:2025zgx}, Pantheon+~\cite{Scolnic:2021amr,Brout:2022vxf}, compressed CMB~\cite{Zhai:2018vmm}, and growth-rate data~\cite{Euclid:2025bxg} prefer specific functional forms for the nonminimal coupling \(F(\varphi)\) and scalar potential \(U(\varphi)\), and whether these reconstructed forms can reproduce or not the DESI-motivated phantom-crossing behavior in the past while remaining consistent with cosmological observations and Solar System tests of gravity.

The manuscript is organized as follows. In Sec.~\ref{sec:theory} we present the scalar-tensor framework in the Jordan frame, derive the background equations and the effective dark-energy description, and motivate our focus on the unscreened branch of the theory. Section~\ref{sec:data_likelihood} summarizes the cosmological datasets and the likelihood used to evaluate each candidate theory, while Sec.~\ref{sec:ga} introduces the GA reconstruction pipeline. Our results for the reconstructed scalar-tensor functions and their cosmological implications are presented in Sec.~\ref{sec:results}. Finally, Sec.~\ref{sec:conclusion} summarizes our conclusions and discusses possible future directions. Further details on the GA pipeline and on the datasets and Solar System constraints employed are provided in Appendixes~\ref{app:implementation_GA} and~\ref{app:datasets}, respectively.

\section{Theoretical Framework}\label{sec:theory}

\subsection{Scalar-tensor Lagrangian in the Jordan frame}

We consider the scalar-tensor action
\begin{equation}\label{eq:action}
    S =
    \int {\rm d}^4x\, \sqrt{-g} \left[ \frac{1}{2}F(\varphi)R - \frac{1}{2}Z(\varphi) \partial^\mu \varphi \partial_\mu \varphi - U(\varphi) \right] + S_{\rm M}
\end{equation}
in units where \(8\pi G_\mathrm{N}=c=1\), with \(G_\mathrm{N}\) the bare gravitational coupling. Here, \(\varphi\) is the scalar field, \(F(\varphi)\) its nonminimal coupling to the Ricci scalar \(R\), \(Z(\varphi)\) controls its kinetic term, \(U(\varphi)\) is the scalar potential, and \(S_{\rm M}\) is the action of the standard matter fields. For \(F(\varphi) = 1\), \(Z(\varphi) = 1\), and \(U(\varphi) = \Lambda\), the GR action with a cosmological constant is recovered.

The action~\eqref{eq:action} is written in the Jordan frame, where matter couples minimally to the metric \(g_{\mu\nu}\), i.e.~\(S_{\rm M}\) does not depend on \(\varphi\), so that the weak equivalence principle is preserved.%
\footnote{The weak equivalence principle states that test bodies fall independently of their composition, while the strong equivalence principle extends this to self-gravitating bodies. Theories satisfying the latter are commonly classified as dark-energy models, and those violating it, such as scalar-tensor theories, as modified gravity~\cite{Joyce:2016vqv}.}
A conformal transformation to the Einstein frame removes the nonminimal coupling to gravity, but induces an explicit coupling between the scalar field and matter. Here we opt for working in the Jordan frame, where matter follows the geodesics of \(g_{\mu\nu}\) and the comparison with observations is more direct.

Any \(Z(\varphi) > 0\) can be absorbed into a redefinition of the scalar field, so we set \(Z(\varphi) = 1\) hereafter; this also fixes the normalization of \(\varphi\), so that the reconstructed functions \(F(\varphi)\) and \(U(\varphi)\) are uniquely defined. We further require \(F(\varphi) > 0\), which ensures a positive effective gravitational coupling and, for \(Z = 1\), the absence of ghosts in both the scalar and tensor sectors~\cite{Esposito-Farese:2000pbo}. Under these conditions, the potential \(U(\varphi)\) drives the late-time acceleration, while the nonminimal coupling \(F(\varphi)\) allows for a richer phenomenology than minimally coupled quintessence, including the possibility of an effective crossing of the phantom divide~\cite{Boisseau:2000pr, Esposito-Farese:2000pbo, Perivolaropoulos:2005yv, Gannouji:2006jm}.

\subsection{\label{sec:background}Background field equations}

Varying the action~\eqref{eq:action} with respect to the metric and the scalar field yields the field equations derived in Ref.~\cite{Esposito-Farese:2000pbo}. We assume a spatially flat Friedmann-Lema\^{\i}tre-Robertson-Walker background,
\begin{equation}
    {\rm d}s^2 =
    -{\rm d}t^2 + a(t)^2 {\rm d}x_i {\rm d}x^i, \label{eq:FRWmetric}
\end{equation}
where \(t\) is the cosmological time, \(x^i\) are comoving Cartesian coordinates, and \(a(t)\) is the scale factor, which we normalize to \(a(t_0) = 1\) (a subscript \(0\) denotes present-day values throughout). For this choice of the metric, the field equations reduce to~\cite{Esposito-Farese:2000pbo}
\begin{subequations}\label{eq:friedmann}
\begin{align}
    3FH^2 &=
    \rho + \frac{1}{2}\dot{\varphi}^{2} - 3H\dot{F} + U, \label{eq:friedmann1}
    \\
    -2F\dot{H} &=
    \rho + p + \dot{\varphi}^{2} + \ddot{F} - H\dot{F}, \label{eq:friedmann2}
\end{align}
\end{subequations}
where \(\rho\) and \(p\) denote the total energy density and pressure of baryons, cold dark matter, and radiation,%
\footnote{We treat these components as non-interacting perfect fluids. Since matter couples minimally to \(g_{\mu\nu}\), the stress-energy tensor of each component is covariantly conserved, so that \(\rho_{\rm b}, \rho_{\rm dm} \propto a^{-3}\) and \(\rho_{\rm r} \propto a^{-4}\).}
an overdot denotes a derivative with respect to \(t\), and \(H \equiv \dot{a}/a\) is the Hubble parameter. On the other hand, the scalar-field equation becomes~\cite{Esposito-Farese:2000pbo}
\begin{equation}\label{eq:KGcosmic}
    \ddot{\varphi} + 3H\dot{\varphi} + U_{\!,\varphi} - 3(2H^2 + \dot{H})F_{\!,\varphi} =
    0,
\end{equation}
where we introduced \(F_{\!,\varphi} \equiv {\rm d}F/{\rm d}\varphi\) and \(U_{\!,\varphi} \equiv {\rm d}U/{\rm d}\varphi\). Together, Eqs.~\eqref{eq:friedmann}--\eqref{eq:KGcosmic} form the closed system that determines the background cosmological evolution once the functions \(F(\varphi)\) and \(U(\varphi)\) are specified. To ensure physical cosmological solutions are found, the basic viability conditions
\begin{equation}
    F > 0
    \qquad{\rm and}\qquad
    H^2 > 0
    \label{eq:viability_conditions}
\end{equation}
are imposed throughout integration of the equations, i.e.~a positive gravitational coupling, and a well-defined expansion history. Details about the numerical implementation of these equations are provided in Appendix~\ref{app:implementation_GA}.

It is possible to reformulate the Friedmann equations~\eqref{eq:friedmann1}--\eqref{eq:friedmann2} in terms of an effective dark-energy fluid description~\cite{Torres:2002pe, Bellini:2019syt, Nesseris:2022hhc}. Denoting by \(\rho_{\rm DE}\) and \(p_{\rm DE}\) the energy density and pressure of the effective component, equivalence between Eqs.~\eqref{eq:friedmann1}--\eqref{eq:friedmann2} and the standard formulation of the Friedmann equations,
\begin{equation}
    3H^2 = \rho + \rho_{\rm DE},
    \qquad
    2\dot{H} + 3H^2 = -(p + p_{\rm DE}),
\end{equation}
is found for
\begin{align}
    \rho_{\rm DE} =&
    \frac{1}{2} \dot{\varphi}^2 + U - 3H\dot{F} + 3H^2(1 - F), \label{eq:effective_energy_density} \\
    p_{\rm DE} =&
    \frac{1}{2} \dot{\varphi}^2 - U + 2H\dot{F} + \ddot{F} - (2\dot{H} + 3H^2)(1 - F). \label{eq:effective_pressure}
\end{align}
Using these definitions, this effective fluid satisfies the standard continuity equation, \(\dot{\rho}_{\rm DE} + 3H(\rho_{\rm DE} + p_{\rm DE}) = 0\), and so the effective equation-of-state parameter, defined as
\begin{equation}
    w_{\rm DE} =
    p_{\rm DE}/\rho_{\rm DE}, \label{eq:effective_eos}
\end{equation}
has the standard interpretation. We note that the definitions from Eqs.~\eqref{eq:effective_energy_density}--\eqref{eq:effective_pressure} used to describe the effective dark energy are not unique (see e.g.~Refs.~\cite{Gannouji:2006jm, Bellini:2014fua}).

The nonminimal coupling also mediates a fifth force. In the quasi-static limit, the growth of matter perturbations is governed  by an effective gravitational coupling \(G_{\rm eff}\), which in general depends on both time and the wave number of the modes. For a scalar field that is light on the scales of interest, in the quasi-static subhorizon regime this coupling becomes scale independent~\cite{Boisseau:2000pr, Tsujikawa:2007gd},
\begin{equation}\label{eq:Geff}
    \mu_G \equiv \frac{G_{\rm eff}}{G_{\rm N}} =
    \frac{1}{F}\,
    \frac{2F + 4F_{\!,\varphi}^2}
     {2F + 3F_{\!,\varphi}^2}.
\end{equation}
In the absence of a screening mechanism that suppresses the fifth force in high-density environments, the scalar field also modifies gravity on small scales, changing the post-Newtonian parameters and the local value of \(G_{\rm eff}\) probed in the Solar System and in SN explosions~\cite{Riazuelo:2001mg}. Although screening may evade these constraints~\cite{Brax:2021wcv, Sirera:2026klo} (e.g.~symmetron~\cite{Hinterbichler:2010es}, chameleon~\cite{Khoury:2003rn, Burrage:2017qrf}, Vainshtein~\cite{VAINSHTEIN1972393}, or k-mouflage~\cite{Babichev:2009ee, Brax:2014wla} screening), all of these mechanisms have their limitations (see Ref.~\cite{Garcia-Garcia:2026nzy} for a review). We therefore opt to remain agnostic about these safeguards, and instead investigate scalar-tensor theories in the absence of screening mechanisms.

\section{\label{sec:data_likelihood} Cosmological datasets}

For each candidate scalar-tensor theory generated by the GA pipeline (see Sec.~\ref{sec:ga} below), we solve the background system of equations~\eqref{eq:friedmann1}--\eqref{eq:friedmann2} and~\eqref{eq:KGcosmic} and compute the corresponding cosmological observables. The reconstructed expansion history \(H(z)\) is used to evaluate distance-based probes, while the effective gravitational coupling \(\mu_G(z)\) enters both the growth-rate equation and the SN luminosity. We constrain the candidates using DESI DR2 BAO measurements~\cite{DESI:2025zgx}, the Pantheon\(+\) SN compilation combined with SH0ES' calibrated Cepheids~\cite{Scolnic:2021amr, Riess:2021jrx, Brout:2022vxf}, compressed CMB distance priors~\cite{Zhai:2018vmm}, and redshift-space distortion measurements of \(f\sigma_8\)~\cite{Euclid:2025bxg}. For the CMB compression, we have checked a posteriori that the reconstructed scalar-tensor models remain sufficiently close to GR at recombination, so that the standard pre-recombination calibration of the sound horizon is valid~\cite{Pedrotti:2026ncd}. Altogether, the vector of cosmological parameters that we let vary freely is
\begin{equation}
    \boldsymbol{\theta} =
    (\Omega_{\rm m0}, H_0, M_{\rm B0}, \Omega_{\rm b0}, \sigma_8),
    \label{eq:vector_cosmoparams}
\end{equation}
where \(\Omega_{\rm m0}\) is the fractional matter density, \(M_{\rm B0}\) is the present-day absolute magnitude of SN, $\Omega_{\rm b0}$ is the fractional baryon density, and $\sigma_8$ is the root-mean square fluctuation of linear matter density field within a comoving sphere of radius $8h^{-1}\, {\rm Mpc}$, where $h = H_0/(100\, {\rm km}\, {\rm s}^{-1}\, {\rm Mpc}^{-1})$. We defined the present-day fractional densities as \(\Omega_{i0} = 8\pi G_{\rm N}\rho_{i0}/(3H_0^2)\), i.e.~using the bare gravitational coupling \(G_{\rm N}\) from the action rather than the present-day value of \(G_{\rm eff}\).

Assuming Gaussian errors, the contribution of each dataset is written as
\begin{equation}
    \chi_i^2 = 
    \Delta\boldsymbol{x}_i^{T} \cdot \boldsymbol{C}_i^{-1} \cdot\Delta\boldsymbol{x}_i,
\end{equation}
where \(\Delta\boldsymbol{x}_i\) is the residual vector between the theoretical prediction and the observed data vector, and \(\boldsymbol{C}_i\) is the corresponding observational covariance matrix. The total objective function minimized by the GA is
\begin{equation}
    \chi^2_{\rm tot} =
    \chi^2_{\rm BAO} + \chi^2_{\rm CMB} + \chi^2_{\rm SN} + \chi^2_{\rm growth} + \chi^2_{\rm local},
\label{eq:total_chi2_main}
\end{equation}
where \(\chi^2_{\rm local}\) incorporates the necessary local gravity constraints to avoid invoking screening mechanisms while remaining compatible with observations. The definitions of the relevant observables, covariance matrices, the supernovae absolute magnitude correction, matter-density perturbation growth equation, and local constraints are collected in Appendix~\ref{app:datasets}.

\section{\label{sec:ga}Genetic Algorithms}

\begin{figure}[t]
    \centering
    \begin{overpic}[abs,unit=1pt]{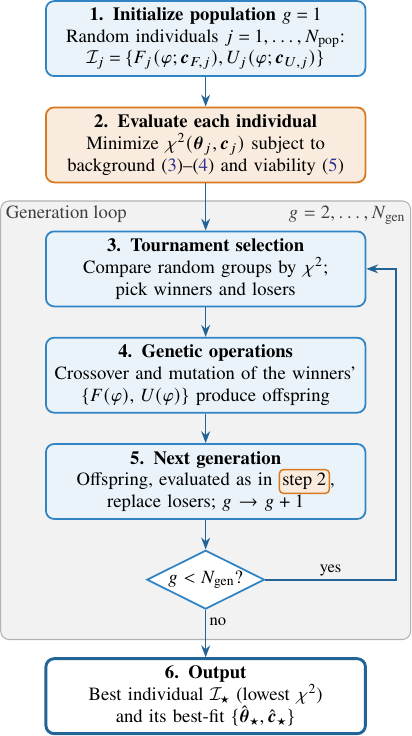}
        \figeqlink{80.65}{271.75}{eq:friedmann}{3}%
        \figeqlink{95.65}{271.75}{eq:KGcosmic}{4}%
        \figeqlink{158.14}{271.75}{eq:viability_conditions}{5}%
        \figsteptarget{21.59}{302.87}{ga:step2}%
        \figsteplink{134.14}{116.08}{24.81}{12.2}{ga:step2}%
    \end{overpic}
    \caption{Flow chart of the genetic-algorithm pipeline. Each individual \(\mathcal{I}_j\) is a pair of functions \(\{F_j(\varphi; \boldsymbol{c}_{F,j}), U_j(\varphi; \boldsymbol{c}_{U,j})\}\) of the scalar field \(\varphi\) with free coefficients \(\boldsymbol{c}_j= (\boldsymbol{c}_{F,j}, \boldsymbol{c}_{U,j})\). Once the \(N_{\rm pop}\) random individuals are generated, each one is evaluated: its cosmological parameters \(\boldsymbol{\theta}_j\) and coefficients \(\boldsymbol{c}_j\) are fitted to the data by minimizing the \(\chi^2\), integrating the background cosmology~\eqref{eq:friedmann}--\eqref{eq:KGcosmic} at each trial point and requiring the viability conditions~\eqref{eq:viability_conditions} (non-viable individuals are assigned \(\chi^2 = \infty\)). This yields the best-fit parameters \(\{\hat{\boldsymbol{\theta}}_j, \hat{\boldsymbol{c}}_j\}\). In every subsequent generation \(g\) (gray panel), tournament selection compares individuals by \(\chi^2\); the winners undergo crossover and mutation to produce offspring, which are evaluated in the same way and replace the losers, while the remaining individuals pass unchanged to the next generation. After \(N_{\rm gen}\) generations, the individual with the lowest \(\chi^2\), \(\mathcal{I}_\star\), is returned together with its best-fit parameters \(\{\hat{\boldsymbol{\theta}}_\star, \hat{\boldsymbol{c}}_\star\}\).}\label{fig:ga_pipeline}
\end{figure}

We employ the GA to propose candidates for the fundamental functions \(F(\varphi)\) and \(U(\varphi)\) that determine the scalar-tensor action~\eqref{eq:action}. Each individual in the population represents a candidate pair,
\begin{equation}
    \mathcal{I}_j =
    \left\{F_j(\varphi), U_j(\varphi)\right\},
\end{equation}
which is constructed from a predefined grammar of elementary functions and algebraic operations that defines the symbolic search space explored by the algorithm. The flow chart in Fig.~\ref{fig:ga_pipeline} illustrates the main steps of the pipeline. Starting from an initial population of \(N_{\rm pop}\) randomly generated individuals, each candidate is evaluated by solving the system of equations~\eqref{eq:friedmann}--\eqref{eq:KGcosmic}. After imposing the viability conditions~\eqref{eq:viability_conditions} and selecting the physical candidates, the corresponding fitness function is computed, which we adopt it to be the \(\chi^2\) of the cosmological data (see Sec.~\ref{sec:data_likelihood}). The optimal values of the cosmological parameters \(\hat{\boldsymbol{\theta}}_j\) and the set of coefficients \(\hat{\boldsymbol{c}}_j\) of an individual are obtained by minimizing the \(\chi^2\) in each case; candidates violating the viability conditions~\eqref{eq:viability_conditions} are assigned \(\chi^2 = \infty\).

After the fitness evaluation, the best-performing individuals are selected through tournament selection. The operations of crossover and mutation are then applied sequentially, each with a given probability: crossover combines the symbolic structure of two parent individuals to generate new offspring, while mutation randomly modifies part of an individual, for instance by replacing a grammar element, or altering an operation. The newly generated offspring replace lower-performing individuals, and the procedure is repeated for successive \(N_{\rm gen}\) generations. Finally, we extract the individual with the lowest $\chi^2$, \(\mathcal{I}_\star = \{F_\star(\varphi), U_\star(\varphi)\}\), together with the optimal cosmological parameters \(\hat{\boldsymbol{\theta}}_\star\) and coefficients of the functions \(\hat{\boldsymbol{c}}_\star\). We refer the reader to Appendix~\ref{app:implementation_GA} for further details.

One caveat of GA resides in their sensitivity to the choice of the hyperparameters and intrinsic stochasticity; even fixed hyperparameters may yield different best candidates \(\mathcal{I}_\star\) when considering different random seeds. This can be alleviated by running \(N_{\rm conf}\) different hyperparameter choices, obtaining an ensemble of best-fit functions \(\{ \mathcal{I}_{\star,k} \}_{k=1}^{N_{\rm conf}}\); here we consider \(N_{\rm conf} = 64\) (see Appendix~\ref{app:implementation_GA} for details). The candidate with the absolute minimum \(\chi_{\rm min}^2\) can then be regarded as the overall set of best-fit functions. Nonetheless, extracting conclusions from this single candidate can be misleading, as it might not capture the overall trend displayed by the set of \(N_{\rm conf}\) best candidates. For this reason, we represent the observables of interest as a weighted average, following a similar prescription as that implemented in the \texttt{GAME} pipeline from Ref.~\cite{Peronaci:2026iur}. Hereafter, we refer to it as \texttt{GAME}-style ensemble. We first define the \(N_{\rm conf}\) normalized weights \(\omega_k\) for the \(k\)th best candidate as an exponential cutoff based on their scored \(\chi_k^2\),
\begin{equation}
    \omega_k =
    \frac{\exp \left[ -\frac{1}{2} \left( \chi_k^2 - \chi_{\rm min}^2 \right) \right]}{\sum_{i=1}^{N_{\rm conf}} \exp \left[ -\frac{1}{2} \left( \chi_i^2 - \chi_{\rm min}^2 \right) \right]}
    ,\qquad
    \sum_{k=1}^{N_{\rm conf}} \omega_k = 1.
\end{equation}
Then, we compute the average of a derived quantity \(f(z)\), such as \(\mu_G(z)\) or the effective \(w(z)\), from the values \(f_k(z)\) obtained with the best-fit functions of each configuration \(k\):
\begin{equation}
    f_\texttt{GAME}(z) =
    \sum_{k = 1}^{N_{\rm conf}} \omega_k f_k(z). \label{eq:f_game}
\end{equation}
Similarly, we can define the corresponding weighted ensemble variance at each redshift to encode the configuration spread between each GA run of the observable \(f(z)\) as%
\footnote{Ref.~\cite{Peronaci:2026iur} instead weights the deviations in Eq.~\eqref{eq:sigma_game} by \(\omega_k^2\) rather than \(\omega_k\). Their variance estimates the uncertainty of the weighted average itself instead of the spread of the ensemble around it.}
\begin{equation}
    \sigma_{f,\texttt{GAME}}^2(z) =
    \sum_{k=1}^{N_{\rm conf}} \omega_k \big( f_{k}(z) - f_\texttt{GAME}(z) \big)^2. \label{eq:sigma_game}
\end{equation}
The resulting $f_\texttt{GAME}(z) \pm \sigma_{f,\texttt{GAME}}$ therefore represents the \texttt{GAME}-style ensemble that accounts for the overall trend of the $N_{\rm conf}$ GA runs and their corresponding spread.

We remark that, due to limited computational resources, we do not perform an additional Bayesian inference over each of the resulting cosmologies. Thus, all \texttt{GAME}-style results should be understood as evaluated over the best-fit cosmological parameters and function coefficients, thereby underestimating the total uncertainty.

\section{\label{sec:results} Results and discussion}

We now present the results of our symbolic search, quoting the overall best-performing scalar-tensor theory and comparing it with the standard flat \(\Lambda{\rm CDM}\) model and the CPL parametrization with equation-of-state parameter \(w(z) = w_0 + w_a z/(1+z)\). We first show the performance of this scalar-tensor theory in terms of the \(\chi^2\) scored, and discuss the constraints obtained for the cosmological parameters. Then, we investigate the cosmological implications resulting from the overall behavior of the GA search by employing the \texttt{GAME}-style weighted average. For the remainder, our reported results will refer to the rescaled potential \(U = U_{\rm phys} / H_0^2\).

\subsection{\label{sec:reconstructed_lagrangian} Reconstructed best scalar-tensor Lagrangian}

Selecting the overall best individual among the \(64\) GA runs we performed, we obtain the following pair of nonminimal coupling and potential functions:
\begin{align}
    F(\varphi) &=
    0.88234 \varphi^2 + e^\varphi(1 - 0.994625 \varphi), \label{eq:best_F} \\
    U(\varphi) &=
    U_0 - 0.636\varphi^2 - \varphi^3. \label{eq:best_U}
\end{align}
As discussed in Appendix~\ref{app:implementation_GA}, the value of \(U_0\) is determined by demanding \(H(z=0) / H_0 = 1\), and so its specific numerical value is straightforward to compute and does not provide further insight. The values of the remaining coefficients in \(F(\varphi)\) and \(U(\varphi)\) are the ones that yield the best fit to the data.

Table~\ref{tab:chi2} contains the \(\chi^2\) values of all the datasets for the \(\Lambda{\rm CDM}\) model, CPL, and the scalar-tensor theory from Eqs.~\eqref{eq:best_F}--\eqref{eq:best_U}, all evaluated at their best-fit cosmological parameters to all data. The symbolic reconstruction improves the fit relative to both models, with \(\Delta\chi^2=-16.6\) with respect to \(\Lambda{\rm CDM}\) and \(-6.0\) with respect to CPL. The largest contribution comes from the Pantheon\(+\)\&SH0ES data. As we discuss below (Sec.~\ref{sec:effective_grav_coupling}), the reconstruction allows the effective gravitational coupling to vary by a few percent over \(0<z\lesssim2.5\), so that the SN absolute magnitude \(M\) evolves with redshift and better fits the observed SN apparent magnitudes. Within \(\Lambda\)CDM, the SH0ES calibration is in tension with the CMB-anchored distance scale, as reflected in the larger CMB \(\chi^2\) of \(\Lambda\)CDM in Table~\ref{tab:chi2}. Part of the improvement therefore reflects the ability of an evolving \(G_{\rm eff}\) to reconcile the two calibrations of \(M\), in the spirit of evolving-\(M\) proposals \cite{Kazantzidis:2020tko,Marra:2021fvf}. The direction of this evolution also relies on the luminosity scaling adopted in Eq.~\eqref{eq:sncorr}. 

The reconstruction also fits the DESI DR2 BAO data better than CPL, though slightly worse than \(\Lambda{\rm CDM}\), and fits the compressed CMB distance priors better than both. It is compatible with the Solar System constraints from the Cassini measurement of the parametrized post-Newtonian (PPN) parameter \(\gamma\) and with lunar laser ranging constraints on the time variation of the effective gravitational coupling, \(\dot{G}_{\rm eff}/G_{\rm eff}\). Despite the \(\sim\!2\sigma\) deviation from the BBN constraint~\cite{Alvey:2019ctk}, the reconstructed scalar-tensor theory stands as a competitive candidate without overly complex functional forms for \(F(\varphi)\) and \(U(\varphi)\).

\begin{table}[t]
    \caption{$\chi^2$ of each dataset at the best fit of $\Lambda$CDM, CPL and the best-performing scalar-tensor theory reported in Eqs.~\eqref{eq:best_F}--\eqref{eq:best_U}, quoted as GA. The last row shows \(\Delta \chi^2 \equiv \chi_{\rm min}^2 - \chi_{\rm min,\Lambda CDM}^2\).} \label{tab:chi2}
    \begin{ruledtabular}
        \begin{tabular}{lrrr}
         & $\Lambda$CDM & CPL & GA \\
        \hline
        Pantheon+\&SH0ES                     & 1551.95 & 1544.43 & 1537.95 \\
        DESI DR2 BAO                         &   10.64 &   14.19 &   12.13 \\
        Planck (compressed)                  &    7.47 &    0.91 &    0.09 \\
        RSD $f\sigma_8$                      &   17.08 &   16.99 &   16.40 \\
        Cassini $\gamma$                     &    0.83 &    0.83 &    0.83 \\
        LLR $\dot{G}_{\rm eff}/G_{\rm eff}$  &    0.27 &    0.27 &    0.00 \\
        BBN $G_{\rm eff}$                    &    0.11 &    0.11 &    4.34 \\
        \hline
        Total \(\chi_{\rm min}^2\)           & 1588.36 & 1577.74 & 1571.74 \\
        \(\Delta \chi^2\)                    &    0.00 &  -10.62 &  -16.62
        \end{tabular}
    \end{ruledtabular}
\end{table}

Despite the improvement in the \(\chi^2\) provided by the reconstructed scalar-tensor theory, we do not interpret it in terms of any model-selection criteria such as the AIC or BIC. The reason is that there is no universal notion of model parameters for these symbolic regression techniques. Besides the number of free coefficients entering the functions, the information carried by their structural complexity should also be accounted for. Since both the definition of this complexity and the identification of the free coefficients as model parameters are intrinsically arbitrary, we simply report the \(\Delta \chi^2\) as a goodness-of-fit quantity without attaching to it any statistical significance over the competing models.

Table~\ref{tab:cosmoparams} contains the cosmological parameters of the \(\Lambda{\rm CDM}\) model, the CPL model, and the best individual found by the GA. For \(\Lambda{\rm CDM}\) and CPL, we quote posterior medians with 68\% credible intervals, obtained with \texttt{Nautilus}~\cite{Lange:2023ydq}, whereas the scalar-tensor constraints are best-fit values obtained for the minimum \(\chi^2\). The matter-density \(\Omega_{\rm m0}\) of the scalar-tensor theory lies slightly below the values of both \(\Lambda{\rm CDM}\) and CPL, and the constraints on the \(\sigma_8\) parameter are compatible for all three models. The Hubble parameter \(H_0\) increases by \(1\, {\rm km}\, {\rm s}^{-1}\, {\rm Mpc}^{-1}\) with respect to both \(\Lambda{\rm CDM}\) and CPL, thereby slightly alleviating the tension between the early- and late-time inferred values of \(H_0\). This larger value causes an increase in the present-day absolute magnitude \(M_{\rm B0}\) value, and a reduction in \(\Omega_{\rm b0}\) due to the CMB anchor on the value for \(\Omega_{\rm b0} h^2\).

\begin{table}[t]
    \caption{Constraints on cosmological parameters for \(\Lambda{\rm CDM}\), CPL, and the single best scalar-tensor theory (GA) reported in Eqs.~\eqref{eq:best_F} and~\eqref{eq:best_U}. The Hubble constant \(H_0\) is given in units of \({\rm km}\, {\rm s}^{-1}\, {\rm Mpc}^{-1}\). For \(\Lambda{\rm CDM}\) and CPL, we quote posterior medians with 68\% credible intervals obtained with \texttt{Nautilus}, whereas for the scalar-tensor theory we present the best-fit values corresponding to the minimum \(\chi^2\).}\label{tab:cosmoparams}
    \begin{ruledtabular}
    \begin{tabular}{lccc}
        Parameter           & \(\Lambda{\rm CDM}\)              & CPL                               & GA \\
        \hline
        \(\Omega_{\rm m0}\) & \(0.2964^{+0.0033}_{-0.0034}\)    & \(0.2994^{+0.0049}_{-0.0050}\)    & \(0.2934\) \\[.25em]
        \(H_0\)             & \(68.85^{+0.28}_{-0.27}\)         & \(68.95^{+0.54}_{-0.53}\)         & \(69.86\) \\[.25em]
        \(M_{\rm B0}\)      & \(-19.398^{+0.008}_{-0.008}\)     & \(-19.382^{+0.013}_{-0.013}\)     & \(-19.361\) \\[.25em]
        \(\Omega_{\rm b0}\) & \(0.04779^{+0.00030}_{-0.00032}\) & \(0.04727^{+0.00076}_{-0.00076}\) & \(0.04590\) \\[.25em]
        \(\sigma_8\)        & \(0.801^{+0.026}_{-0.027}\)       & \(0.795^{+0.026}_{-0.026}\)       & \(0.806\) \\[.25em]
        \(w_0\)             & \(-1\)                            & \(-0.873^{+0.055}_{-0.054}\)      & --- \\[.25em]
        \(w_a\)             & \(0\)                             & \(-0.64^{+0.22}_{-0.23}\)         & --- \\
    \end{tabular}
    \end{ruledtabular}
\end{table}

\subsection{\texorpdfstring{Generational evolution of the minimum \(\chi^2\)}{Generational evolution of the minimum χ²}}

\begin{figure}[t]
    \centering
    \includegraphics{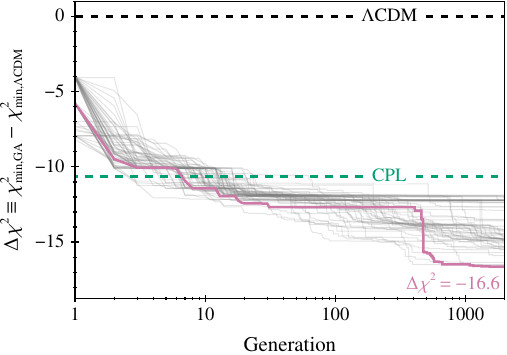}
    \caption{Generational evolution of \(\chi_{\rm min}^2\) for the \(64\) independent runs, relative to \(\Lambda{\rm CDM}\). The pink line highlights the run which scored the minimum \(\chi^2\) out of all runs after termination; gray lines illustrate all other realizations. The green dashed line marks the CPL \(\chi_{\rm min}^2\) value.}
    \label{fig:chi2_independent_runs}
\end{figure}

Figure~\ref{fig:chi2_independent_runs} shows the generational evolution of \(\chi_{\rm min}^2\) of each independent run, relative to the \(\Lambda\)CDM value of \(\chi_{\rm min}^2\). The solid pink line highlights the run that reached the lowest \(\chi^2\) at termination, which yields the best \(F(\varphi)\) and \(U(\varphi)\) functions, while the thin gray lines show all other independent realizations. All runs outperform both \(\Lambda\)CDM and CPL, showing that the symbolic search consistently finds competitive solutions. The final values of \(\chi_{\rm min}^2\) differ by \(\sim\!5\) units among the runs, reflecting the stochastic nature of the algorithm, while the qualitative behavior of the reconstructed quantities is shared across runs, as reflected in the \texttt{GAME}-style spread. As is generally the case for GAs, which lack a formal convergence criterion, longer runs could yield further modest reductions in \(\chi_{\rm min}^2\), at a computational cost beyond the scope of this work.

\subsection{Cosmological implications}

\subsubsection{\label{sec:effective_grav_coupling}Effective gravitational coupling}

\begin{figure}[t]
    \centering
    \includegraphics{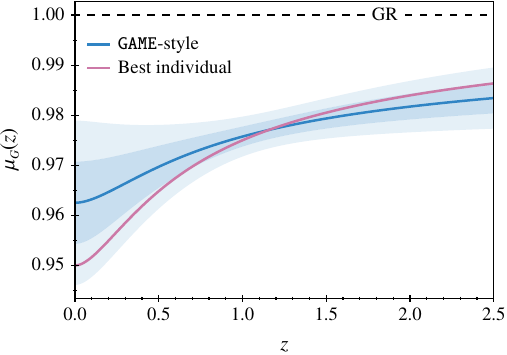}
    \caption{Evolution of the effective gravitational coupling for the weighted average across the GA outputs (solid blue line, \texttt{GAME}-style) and for the best scalar-tensor theory given by the \(F(\varphi)\) and \(U(\varphi)\) functions in Eqs.~\eqref{eq:best_F} and~\eqref{eq:best_U} (solid pink line). The dark- and light-blue shaded regions correspond to the \(1\sigma\) and \(2\sigma\) weighted ensemble variance of the GA. The horizontal black dashed line marks the standard gravitational coupling limit \(G_{\rm eff} (z) = G_{\rm N}\).}\label{fig:effective_gravitational_coupling}
\end{figure}

Figure~\ref{fig:effective_gravitational_coupling} shows the evolution of the effective gravitational coupling. The blue line is the \texttt{GAME}-style weighted average computed from the ensemble of scalar-tensor configurations found by the GA, and the dark- and light-shaded bands are the weighted ensemble variances corresponding to \(1\sigma_\texttt{GAME}\) and \(2\sigma_ \texttt{GAME}\); the pink solid line represents the inferred effective gravitational coupling from the best functions given by Eqs.~\eqref{eq:best_F}--\eqref{eq:best_U}. The coupling increases with redshift in both cases, in qualitative agreement with the reconstruction of Ref.~\cite{Efstratiou:2025iqi}. Over the redshift range \(0 < z < 2.5\), it varies by \(\sim\!2\)--\(4\%\), which translates into an evolution of the SN absolute magnitude of approximately \(0.035\)--\(0.06\, {\rm mag}\) and provides the additional freedom in the SN fit discussed above. At high redshift, \(\mu_G \to 1\) as imposed by the initial conditions of the field (see Appendix~\ref{app:implementation_GA}), so that GR is recovered by recombination. The variation of the coupling between Big Bang nucleosynthesis (BBN) and the present day lies within the \(\sim\! 2\sigma\) bound of Ref.~\cite{Alvey:2019ctk}, which we included in the likelihood to avoid larger tensions with this measurement.

We note that \(\mu_G(z=0) \neq 1\) does not conflict with laboratory measurements of Newton's constant, as Cavendish-type experiments measure the present-day effective coupling \(G_{\rm eff}(z=0)\) rather than the bare constant \(G_{\rm N}\) appearing in the action. Local tests instead constrain the PPN parameter \(\gamma\) and the present-day time variation \(\dot{G}_{\rm eff}/G_{\rm eff}\), both of which are included in our analysis and satisfied by our reconstructions.

\subsubsection{Equation of state}

\begin{figure}[t]
    \centering
    \includegraphics{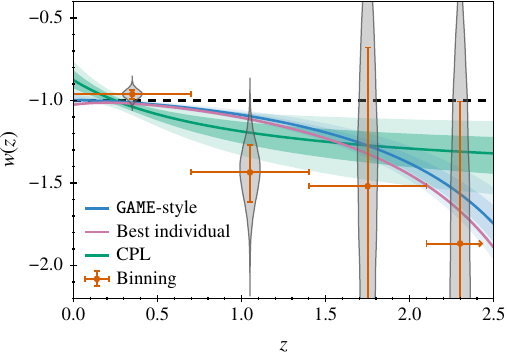}
    \caption{Effective equation-of-state parameter from the genetic-algorithm outputs. The solid blue line shows the \texttt{GAME}-style weighted average, and the dark- and light-blue bands represent the \(1\sigma_\texttt{GAME}\) and \(2\sigma_\texttt{GAME}\) configuration spread, see Eq.~\eqref{eq:sigma_game}. The solid pink line shows the best individual, with \(F(\varphi)\) and \(U(\varphi)\) given by Eqs.~\eqref{eq:best_F}--\eqref{eq:best_U} and evaluated at the best-fit cosmological parameters and coefficients. The solid green line and bands show the median and \(1\sigma\) and \(2\sigma\) regions of the CPL fit to the same data. The orange points show the median and \(1\sigma\) errors of \(w\) in three uniform bins in the range \(0 < z < 2.1\), and an open-ended bin at \(z > 2.1\), with horizontal bars indicating the bin widths, and the gray violins show the one-dimensional posterior of each bin amplitude. The dashed black line marks the phantom divide, \(w = -1\).}\label{fig:equation_of_state}
\end{figure}

Figure~\ref{fig:equation_of_state} displays the effective equation-of-state parameter~\eqref{eq:effective_eos} from the scalar-tensor realizations, with the blue and pink lines denoting the \texttt{GAME}-style weighted average and the best-individual reconstruction from Eqs.~\eqref{eq:best_F}--\eqref{eq:best_U}, respectively. For reference, we also include uniformly binned values of the equation-of-state parameter in the range \(0 < z < 2.1\) and an open-ended bin at \(z > 2.1\) in orange~\cite{DESI:2025fii},%
\footnote{Specifically, we consider the binned function
\begin{equation*}
w(z) =
w_0 + \sum_{i=1}^N \frac{(w_i - w_{i-1})}{2} \left[ 1 + \tanh \left( \frac{z - z_i}{s} \right) \right],
\end{equation*}
where \(w_0, \dots, w_N\) are the amplitudes of the \(N + 1\) bins, \(z_i\) are the bin edges, and \(s\) is the smoothing scale, which we fix to \(s = 0.02\) following Ref.~\cite{DESI:2025fii}. We use \(N = 3\) with edges \(z_i = 0.7, 1.4, 2.1\), and in Fig.~\ref{fig:equation_of_state} we show the last bin amplitude (\(z > 2.1\)) at \(z = 2.3\) for visualization only.}
with horizontal bars denoting the redshift range of each bin, the vertical bars denoting the \(1\sigma\) errors, and the gray violins illustrating the one-dimensional posterior of each bin amplitude. To ensure a consistent comparison, the bin amplitudes are fitted to the same datasets as all other models, rather than adopting the values reported in Ref.~\cite{DESI:2025fii}, which were obtained from a different data combination.

The reconstructions agree qualitatively with the CPL equation-of-state parameter (green line and \(1\sigma\) and \(2\sigma\) bands in Fig.~\ref{fig:equation_of_state}) at \(z \gtrsim 0.5\), where both are phantom. At low redshift, however, they depart from it: CPL crosses into the quintessence regime at \(z \lesssim 0.3\), whereas the best-performing scalar-tensor theory remains phantom over the whole redshift range, and the \texttt{GAME}-style reconstruction is consistent with \(w = -1\) at \(z = 0\) within its spread. Table~\ref{tab:chi2} helps to separate the two effects. Both models obtain a comparable improvement in the compressed CMB, which we associate with their common phantom behavior at intermediate redshifts. They differ in how they fit the SN data. CPL can only do so through the time evolution of \(w(z)\), which drives it to \(w > -1\) at late times and hence across the phantom divide. The scalar-tensor reconstructions instead fit the SN data through the time-varying gravitational coupling (Fig.~\ref{fig:effective_gravitational_coupling}), which lets the SN absolute magnitude evolve with redshift and yields a larger improvement in this sector. The phantom behavior is therefore common to both models, while the crossing is specific to CPL, where it is the only way to accommodate the low-redshift SN data.

These findings are not in conflict with previous studies constraining scalar-tensor theories. Ref.~\cite{Wolf:2025jed} considered a quadratic nonminimal coupling \(F(\varphi) = 1 - \xi \varphi^2\) with a potential \(U(\varphi) = U_0 + \alpha \varphi + m^2 \varphi^2/2\) in a screened scenario, in which the effective gravitational coupling affects neither the SN luminosities nor the growth of matter-density perturbations, and found that the effective equation-of-state parameter crosses the phantom divide. A similar behavior was recently reported in Ref.~\cite{Argudo-Panes:2026qkw} for the unscreened case, using the same coupling with a linear potential and the same Solar System constraints as in this work. There, these constraints could only be satisfied by imposing \(\varphi(z=0) = 0\), which restricts \(G_{\rm eff}\) to sub-percent deviations from \(G_{\rm N}\)~\cite{Argudo-Panes:2026qkw}. In neither case can the effective gravitational coupling improve the fit to the SN data, which must instead be achieved by modifying the background expansion history. This ultimately drives \(w(z)\) to cross the phantom divide. Here, the greater flexibility of our symbolic reconstructions avoids this restriction. The Solar System constraints are satisfied while \(G_{\rm eff}\) deviates by a few percent from \(G_{\rm N}\) today, with BBN becoming the limiting constraint. The data are then better fitted through a time-evolving gravitational coupling rather than a phantom crossing. This result also agrees with the findings of Ref.~\cite{Efstratiou:2025iqi} when combining a CPL background with a parametrization of \(\mu_G(z)\).

We remind the reader that, because of the high computational cost, we have not performed a full Bayesian inference for the scalar-tensor candidates entering the weighted average. Both the \texttt{GAME}-style reconstruction and the best individual are evaluated at their best-fit values, and do not propagate the observational uncertainties in the cosmological parameters or in the coefficients in \(F(\varphi)\) and \(U(\varphi)\). Including these uncertainties would broaden the bands in Fig.~\ref{fig:equation_of_state} beyond the \texttt{GAME}-style spread, and could also shift their medians. The absence of a phantom crossing should therefore not be over-interpreted, since phantom crossing may remain compatible with the data once these uncertainties are taken into account.

\subsubsection{Scalar-tensor functionals}

\begin{figure}[t]
    \centering
    \includegraphics{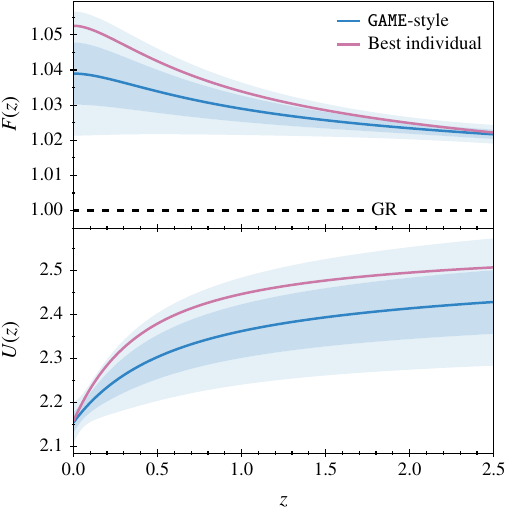}
    \caption{Evolution of the nonminimal coupling \(F(z)\) and the potential \(U(z)\) (in units of \(H_0^2\)). The solid blue line shows the \texttt{GAME}-style weighted average over the best individuals of all GA runs, with dark- and light-blue envelopes denoting the \(1\sigma_{\texttt{GAME}}\) and \(2\sigma_{\texttt{GAME}}\) weighted ensemble spread; the solid pink line shows the overall best candidate, with \(F(\varphi)\) and \(U(\varphi)\) given by Eqs.~\eqref{eq:best_F}--\eqref{eq:best_U}, evaluated along its background solution. The dashed black line in the top panel indicates the GR limit \(F = 1\).}\label{fig:functionals}
\end{figure}

Figure~\ref{fig:functionals} illustrates the redshift evolution of the nonminimal coupling \(F(z)\) (top panel) and the field potential \(U(z)\) (bottom panel), with the blue curve and envelopes showing the \texttt{GAME}-style average and \(1\sigma_\texttt{GAME}\) and \(2\sigma_\texttt{GAME}\) ensemble variances, and the pink curve representing the overall best scalar-tensor theory with functions given by Eqs.~\eqref{eq:best_F}--\eqref{eq:best_U}. The \(F(z)\) reconstruction follows the trend of the effective gravitational coupling, deviating from the GR limit \(F(\varphi) = 1\) at low-redshifts, and approaching \(F(\varphi) \to 1\) at large redshifts. The decreasing tendency of the coupling with redshift is in good agreement with the reconstruction from Ref.~\cite{Efstratiou:2025iqi}, although the analyses differ in the offset of their present-day values.

Regarding the potential \(U(z)\), its value increases with redshift, in contrast to the decreasing trend observed in the reconstruction from Ref.~\cite{Efstratiou:2025iqi}, and tends to a constant value at higher redshifts. The spread in the \texttt{GAME}-style variance is smaller in the present-day, as \(U_0\) is anchored by the constraint \(H(z=0)/H_0 = 1\), which fixes the matter and the effective dark-energy densities to values close to the \(\Lambda{\rm CDM}\) ones. At larger redshifts, \(z \gtrsim 1\), this spread becomes larger as the effects of the scalar field potential on the background evolution become subdominant compared with matter.

\subsubsection{Background expansion history}

\begin{figure}[t]
    \centering
    \includegraphics{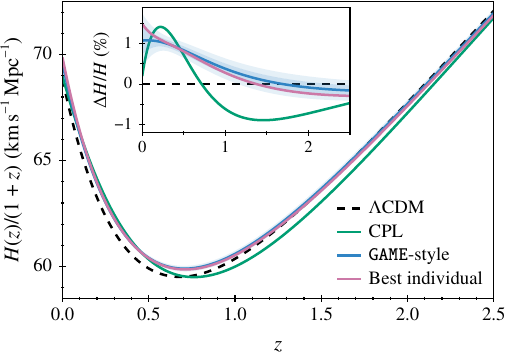}
    \caption{Expansion history for the reference \(\Lambda{\rm CDM}\) and CPL models, as well as for the best-performing scalar-tensor theory with functions given by Eqs.~\eqref{eq:best_F}--\eqref{eq:best_U} and the \texttt{GAME}-style weighted average. All models are evaluated at their best-fit values. Inset figure: expansion history relative to \(\Lambda{\rm CDM}\), with \(\Delta H/H \equiv (H_{\rm model} - H_{\rm \Lambda CDM}) / H_{\rm \Lambda CDM}\).}
    \label{fig:hubble}
\end{figure}

Figure~\ref{fig:hubble} represents the expansion history \(H(z)\) for the \(\Lambda{\rm CDM}\) and the CPL models, as well as the \texttt{GAME}-style weighted average and spread and the best-performing scalar-tensor theory, all evaluated at their best-fit values. We also include an inset figure representing the percent deviation of CPL, the \texttt{GAME}-style average, and the best individual relative to \(\Lambda{\rm CDM}\). We observe that the expansion history predicted by CPL and the reconstructed scalar-tensor theories differs over the redshift range \(0 < z < 2.5\). The phantom crossing predicted by CPL causes the Hubble function to exhibit a dynamic that goes from values above the expansion rate of \(\Lambda{\rm CDM}\), to values that remain below at larger redshifts, when dark energy is phantom. The reconstructed scalar-tensor theories, however, produce an expansion rate that remains larger than \(H_{\Lambda{\rm CDM}}(z)\) until \(z \!\sim\! 1.5\)--\(2\). This is a consequence of the phantom behavior exhibited by the effective dark energy of these models, which naturally yields a higher value of the Hubble constant \(H_0\) than the one observed for \(\Lambda{\rm CDM}\) and CPL~\cite{Lee:2022cyh}.

\section{Conclusions}\label{sec:conclusion}

In this work we have presented a data-driven framework for reconstructing the scalar-tensor Lagrangian directly from cosmological observations. A genetic algorithm evolves symbolic expressions for the nonminimal coupling $F(\varphi)$ and the potential $U(\varphi)$, and each candidate is evaluated through a forward model that determines the expansion history, the effective gravitational coupling and the growth of structure self-consistently. Candidates are ranked against DESI DR2 BAO, Pantheon\(+\) supernovae calibrated with SH0ES Cepheids, compressed CMB distance priors and growth-rate data, while Solar System bounds on the post-Newtonian parameter $\gamma$ and on $\dot{G}_{\rm eff}/G_{\rm eff}$, together with Big Bang nucleosynthesis constraints on the evolution of \(G_{\rm eff}\), restrict the theory in the absence of a screening mechanism.

Across \(64\) independent runs, the best reconstruction improves the fit relative to $\Lambda$CDM by $\Delta\chi^2 = -16.6$, compared with $\Delta\chi^2 = -10.6$ for the CPL parametrization, although this comparison does not account for the freedom of the functional search. The preferred coupling is well described by \(F(\varphi) = 0.88234 \varphi^2 + e^\varphi(1 - 0.994625 \varphi)\), and the \texttt{GAME}-style ensemble reproduces a phantom behavior, in good agreement with the CPL parametrization. Because the Cassini bound forces $F_{,\varphi}$ to be small today, the late-time evolution of $F$ translates almost directly into a change of the locally measured gravitational coupling relative to early times. These results indicate that unscreened scalar-tensor gravity can accommodate the phantom behavior suggested by the current data (but not the crossing suggested by CPL) while remaining consistent with local tests of gravity.

The framework presented here can be extended in several directions. First, the choice of fitness function deserves further study. We rank candidates by their $\chi^2$ alone, which rewards the closest fit to the data regardless of how complex the functions are. This makes the reconstruction prone to overfitting and to poor extrapolation beyond the range covered by the data. It also complicates the comparison with fixed parametrizations such as CPL: because the algorithm explores a large space of functional forms, it can find some improvement in $\chi^2$ even if the underlying cosmology is $\Lambda$CDM, and counting only the free coefficients of the final expressions underestimates this freedom. Information criteria~\cite{Akaike:1974vps,Schwarz:1978xx} penalize the number of parameters but not the search itself, while the information-theoretic approach of Ref.~\cite{Bartlett:2022kyi} also penalizes the structural complexity of each expression. In early tests, however, the latter penalty was so strong that no deviation from $\Lambda$CDM was favored. Designing a fitness function for Lagrangian-level reconstructions that balances goodness of fit against structural complexity remains an open problem. Second, the observational uncertainties are not yet propagated into the reconstructed functions. The \texttt{GAME}-style weighting method used here is built from best-fit cosmologies, so its spread reflects run-to-run variation rather than statistical error. Techniques such as bootstrap Monte Carlo~\cite{Nesseris:2010ep} or the path-integral approach~\cite{Nesseris_2012} exist when a GA fits the data directly, and extending them to reconstructions at the level of the Lagrangian would be worth exploring. Third, we have restricted the analysis to the unscreened branch of the theory, where Solar System bounds force $F_{,\varphi}$ to be small today. Including screening mechanisms would relax this requirement and could allow larger departures from General Relativity.

Upcoming data will sharpen these questions considerably. Further DESI releases, new supernova samples, the growth-rate and weak-lensing measurements from Euclid will test the late-time evolution of the gravitational coupling independently of the background expansion. They will also address a more basic question: whether the apparent crossing of the phantom divide is a genuine feature of the Universe, or an artifact of current datasets and of the parametrizations used to describe them. If it persists, it would point to physics beyond a cosmological constant and, as shown here, scalar-tensor gravity offers a natural setting for it.

\begin{acknowledgments}
The authors would like to thank Savvas Nesseris, Leandros Perivolaropoulos, and Matteo Peronaci for enlightening and helpful discussions. M.~A.~acknowledges financial support from the Basque Government Grant No.~PRE\_2025\_2\_0301, the Egonlabur mobility program No.~EP\_2026\_1\_0083, and the Project PID2025-174372NB-I00 funded by MICIU/AEI/10.13039/501100011033. G.~A.~is supported by the Spanish Research Agency's Consolidaci\'on Investigadora 2024 grant CNS2024-154430. This paper is based upon work from COST Action CA21136 \textit{Addressing observational tensions in cosmology with systematics and fundamental physics} (CosmoVerse), supported by COST (European Cooperation in Science and Technology). 

Computational resources were provided by the Hyperion cluster at the Donostia International Physics Center (DIPC) and the Hydra cluster at the Instituto de F\'isica Te\'orica (IFT, CSIC-UAM). We acknowledge HPC support by Emilio Ambite, staff hired under the Generation D initiative, promoted by Red.es, an organisation attached to the Spanish Ministry for Digital Transformation and the Civil Service, for the attraction and retention of talent through grants and training contracts, financed by the Recovery, Transformation and Resilience Plan through the European Union’s Next Generation funds. The authors also acknowledge the use of AI assistance (GPT Sol 5.6 and Claude Opus 5.5) for improving performance of the numerical codes and refining diagrams and figures for clarity. All content was subsequently reviewed and verified by the authors, who take full responsibility.
\end{acknowledgments}

\section*{Data Availability}
%The codes developed for the analysis will be made publicly available upon publication of this article in a peer-reviewed journal. 
The code and data that support the findings of this manuscript are openly available~\cite{Artola:2026github}.

\appendix

\section{\label{app:implementation_GA} Implementation of the Genetic Algorithm}

In this Appendix we describe the details of the pipeline we skipped in Secs.~\ref{sec:background} and~\ref{sec:ga}. Due to the demanding computational resources required to run the pipeline, our codes are entirely written in \texttt{Julia}~\cite{Julia-2017}, a high-level programming language aimed for high-performance computing. Its efficient differential equation solvers~\cite{DifferentialEquations.jl-2017} and optimizers~\cite{Optim.jl-2018}, together with easy-to-implement symbolic expressions and genetic operators~\cite{cranmer2023interpretablemachinelearningscience} provide the baseline of our pipeline, which we explain below step by step.  All plots were produced using \texttt{Makie.jl}%
\footnote{\url{https://github.com/MakieOrg/Makie.jl}.}%
~\cite{DanischKrumbiegel2021}.

\paragraph{Symbolic representation.}
The symbolic components of an individual \(\mathcal{I}_j = \{F_j(\varphi), U_j(\varphi)\}\) are implemented using \texttt{DynamicExpressions.jl}%
\footnote{\url{https://github.com/SymbolicML/DynamicExpressions.jl}.},
which permits the efficient evaluation of dynamically generated expressions without compiling each candidate separately. Its native interoperability with \texttt{SymbolicRegression.jl}%
\footnote{\url{https://github.com/astroautomata/SymbolicRegression.jl}.}%
~\cite{cranmer2023interpretablemachinelearningscience}
also allows us to adapt its tree-generation and genetic operators to our implementation.

\begin{figure}[t]
    \centering
    \includegraphics{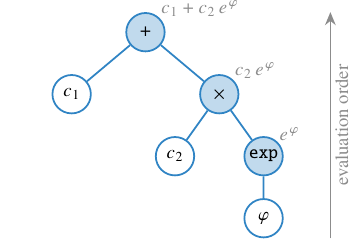}
    \caption{Example of the rooted, ordered tree used to encode the symbolic expression \(c_1 + c_2\,e^{\varphi}\). Operator nodes are shown in filled blue and terminal nodes (leaves) in white. The number of children of each node is fixed by the arity of its symbol: two for the binary operators \(+\) and \(\times\), one for the unary operator \texttt{exp}. Gray labels show the sub-expression evaluated at each operator node. The structural complexity of this expression (i.e.~the number of nodes) is \(k = 6\).}
    \label{fig:tree-encoding}
\end{figure}

More specifically, the symbolic expressions are encoded as rooted, ordered trees. Each node contains a symbol drawn from a predefined grammar, and the arity of that symbol determines the number of child nodes. Nullary (\(0\)-ary) symbols have no children and therefore form leaves, or terminal nodes, of the tree. In the present implementation, the terminal symbols are the scalar field \(\varphi\) and the real-valued coefficients \(c_{i,j}\) (\(i\)th coefficient of the \(j\)th individual). Unary (\(1\)-ary) nodes have one child and here are restricted to the exponential operator \texttt{exp}, whereas binary (\(2\)-ary) nodes have two ordered children and contain one of the algebraic operators \(\{+, -, \times\}\). The root contains the outermost operation, and the corresponding symbolic expression is evaluated recursively from the terminal nodes toward the root. Figure~\ref{fig:tree-encoding} represents the tree-encoding structure of a functional that could be generated by the GA.

We define the structural complexity of an individual as
\begin{equation}
    k_j = 
    k_{F,j} + k_{U_\texttt{GA},j},
\end{equation}
where \(k_{F,j}\) and \(k_{U_\texttt{GA},j}\) denote the total numbers of nodes in the evolved \(F_j\) and \(U_{\texttt{GA},j}\) trees, respectively, including their terminal nodes. The fixed additive structure introduced below through \(U_0\) is common to all candidates and is not included in this count. Each evolved tree is independently constrained to contain at most \(k_{\rm max}\) nodes, and hence \(k_j \leq 2k_{\rm max}\).

\paragraph{Population initialization.}
The GA begins with a population of \(N_{\rm pop}\) randomly generated individuals, followed by \(N_{\rm gen}\) evolutionary updates. The two functions associated with each individual are initialized according to
\begin{equation}
    F_j(\varphi) =
    F_{\texttt{GA},j}(\varphi),
    \qquad
    U_j(\varphi) =
    U_0 + U_{\texttt{GA},j}(\varphi),
\end{equation}
where the subscript \texttt{GA} denotes the symbolic degrees of freedom evolved by the algorithm. The target sizes of the \(F_j(\varphi)\) and \(U_{\texttt{GA},j}(\varphi)\) are drawn independently from the discrete uniform distribution on \(\{1, \ldots, k_{\rm max}\}\); trees with the prescribed number of nodes are then constructed adapting the population initialization method in \texttt{SymbolicRegression.jl}.

\paragraph{Candidate evaluation and parameter optimization.}
Every candidate, whether generated during initialization or as an offspring produced by crossover and mutation, is processed through the same physical and statistical evaluation. For a given vector of cosmological parameters \(\boldsymbol{\theta}\) and coefficients of the symbolic functions \(\boldsymbol{c}_j\), the background system~\eqref{eq:friedmann}--\eqref{eq:KGcosmic} is integrated using the non-stiff \texttt{Tsit5} solver%
\footnote{\url{https://github.com/SciML/OrdinaryDiffEq.jl}.}
over the redshift interval \(z \in [0, 10^9]\) (we reformulated the equations in terms of the e-folds $N = \ln a$ but omit the details here). The initial value of the scalar field at early times \(\varphi_{\rm ini}\) is selected so that \(F(\varphi_{\rm ini}) = 1\),%
\footnote{We use the \texttt{find\_zeros} routine within \texttt{Roots.jl}: \url{https://github.com/JuliaMath/Roots.jl}. More specifically, we search the initial condition in the range \(\varphi_{\rm ini} \in [-1, 1]\) and select the value that is closer to \(\varphi = 0\). If no root is found, the individual is rejected and assigned \(\chi^2 = \infty\).}
whereas the initial value of its derivative \(\varphi_{\rm ini}^\prime\) is set to zero, consistent with an epoch dominated by the Hubble friction. Then, at each evaluation, the additive parameter \(U_0\) is determined through a shooting procedure that imposes the present-day closure relation \(H(z=0)/H_0 = 1\), thereby ensuring that the scalar field evolution is compatible with other cosmological parameters (akin to what is done in \texttt{hi\_class}~\cite{Bellini:2019syt}). The resulting solution is required to satisfy the basic viability conditions~\eqref{eq:viability_conditions}, \(F_j(\varphi) > 0\) and \(H^2 > 0\), throughout the integration; candidates for which the background integration or shooting fails, or for which any of these conditions is violated, are assigned an infinite fitness, \(\chi_j^2 = \infty\), effectively excluding them from selection.

For every viable symbolic structure, the cosmological parameters \(\boldsymbol{\theta}\) and the numerical coefficients \(\boldsymbol{c}_j\) are treated as nuisance parameters and profiled out by minimizing the total observational \(\chi_{{\rm tot},j}^2\) defined in Sec.~\ref{sec:data_likelihood} (see also Appendix~\ref{app:datasets}). More specifically, we determine
\begin{equation}
    \bigl(
        \hat{\boldsymbol{\theta}}_j,
        \hat{\boldsymbol{c}}_j
    \bigr)
    =
    \underset{\boldsymbol{\theta},\,\boldsymbol{c}_j}
    {\operatorname{argmin}}\,
    \chi_{{\rm tot},j}^2
    \bigl(\boldsymbol{\theta},\boldsymbol{c}_j\bigr)
\end{equation}
using the \texttt{Nelder-Mead} optimizer,%
\footnote{\url{https://github.com/JuliaNLSolvers/Optim.jl}. We also impose a wall-time limit of \(2\) seconds of minimization per individual to avoid stagnation of very convoluted expressions, which would otherwise unnecessarily slow down the pipeline. We have checked that this wall-time limit does not affect our results for viable individuals, as each individual has a few hundred background evaluations to get close enough to the minimum \(\chi^2\).}
solving the background and ensuring physical viability in every step. This allows us to rank each scalar-tensor candidate by the best fitness possible for each proposed functional form. Selecting inappropriate coefficients or unphysical cosmological parameters would otherwise unfairly penalize potentially viable functional forms.

\paragraph{Selection and genetic operators.}
Once all members of a generation have been evaluated, the new generation is formed by selecting the best-performing fraction of the population \(r_{\rm sel}\) via a tournament-selection procedure adapted from the \texttt{GATO}%
\footnote{\url{https://github.com/snesseris/Genetic-Algorithms}.}
code (see Refs.~\cite{Bogdanos:2009ib, Nesseris_2012} for further details) and implemented natively in \texttt{Julia}. The remaining population slots are then obtained by choosing random parents from the winners and applying the genetic operators of crossover and mutation with probabilities \(p_{\rm cross}\) and \(p_{\rm mut}\), respectively. These operators are adapted from \texttt{SymbolicRegression.jl} to act on the paired symbolic structure \(\{F_j, U_{\texttt{GA},j}\}\), while enforcing the prescribed grammar and maximum tree size. Every resulting offspring is passed through the same shooting, viability, and parameter-optimization pipeline.

\paragraph{Termination and output.}
Selection, crossover and mutation, and candidate evaluation are repeated for \(N_{\rm gen}\) generations, retaining the lowest-fitness individual encountered up to each generation independently of subsequent population changes. Then, at the end of the run, the algorithm returns the best-performing individual \(\mathcal{I}_\star = \{F_\star(\varphi), U_\star(\varphi)\}\), its optimized cosmological parameters \(\hat{\boldsymbol{\theta}}_\star\), and the optimized numerical coefficients \(\hat{\boldsymbol{c}}_\star\).

\paragraph{Hyperparameters of the GA runs.}
Table~\ref{tab:ga_hyperparameters} summarizes the GA hyperparameters used in this work. Values of the population size \(N_{\rm pop}\) and the number of generations \(N_{\rm gen}\) were selected by balancing the computational cost of evaluating the set of individuals, as well as the amount of generations required to not find significant improvements in the best-performing individuals. Nonetheless, we remind the reader that there is no proper notion of convergence for GAs, and the runs should ideally have the largest number of population size and generations possible.

\begin{table}[t]
    \caption{Hyperparameters and symbolic grammar adopted for the genetic-algorithm runs. The \(4\) different values of the crossover rate \(p_{\rm cross}\), the mutation rate \(p_{\rm mut}\), and the ``random'' seeds \(n_{\rm seed}\) are used to perform \(N_{\rm conf} = 64\) independent runs, one for each combination.}
    \label{tab:ga_hyperparameters}
    \begin{ruledtabular}
    \begin{tabular}{lll}
        Hyperparameter & Symbol & Value \\
        \hline
        Population size & \(N_{\rm pop}\) & \(1000\) \\
        Generations & \(N_{\rm gen}\) & \(2000\) \\
        Crossover rate & \(p_{\rm cross}\) & \(\{0.2, 0.4, 0.6, 0.8\}\) \\
        Mutation rate & \(p_{\rm mut}\) & \(\{0.2, 0.4, 0.6, 0.8\}\) \\
        Selection rate & \(r_{\rm sel}\) & \(0.5\) \\
        Maximum nodes & \(k_{\rm max}\) & \(15\) \\
        ``Random'' seeds & \(n_{\rm seed}\) & \(\{104729, 130363, 181081, 206369\}\) \\
        \hline
        \multicolumn{3}{l}{Grammar} \\
        Nullary symbols & & \(\{\varphi,\;c_i\}\) \\
        Unary operators & & \texttt{exp} \\
        Binary operators & & \(\{+,\;-,\;\times\}\) \\
    \end{tabular}
    \end{ruledtabular}
\end{table}

To mitigate the impact of hyperparameter selection and stochasticity of the GA, we considered four different values for the crossover and mutation rates \(p_{\rm cross}\) and \(p_{\rm mut}\), respectively, as well as four different random initialization seeds as exemplified in Table~\ref{tab:ga_hyperparameters}. More specifically, for each possible combination of these hyperparameters and maintaining the remaining fixed, we performed a total of \(N_{\rm conf} = 64\) independent runs. We selected the best performing individual out of all of them, and used the resulting ensemble to produce the \texttt{GAME}-style plots throughout Sec.~\ref{sec:results}.

\section{\label{app:datasets} Observational datasets}

\paragraph{Baryon acoustic oscillations.}
We employ the latest BAO measurements from the DESI DR2 compilation~\cite{DESI:2025zgx}.%
\footnote{The full precision data points, as well as the covariance matrix, are publicly available at the \texttt{Cobaya} repository: \url{https://github.com/CobayaSampler/bao_data/tree/master/desi_bao_dr2}.}
This catalog provides constraints over seven redshift bins in the range \(0.295 \leq z_{\rm eff} \leq 2.33\) for the comoving distance \(D_{\rm M}/r_{\rm d}\), the Hubble distance \(D_{\rm H}/r_{\rm d}\), and the angle-averaged distance \(D_{\rm V}/r_{\rm d}\), all normalized by the comoving sound horizon at the drag epoch, \(r_{\rm d} \equiv r_{\rm s}(z_{\rm d})\). We compute the sound horizon using its standard definition in terms of the sound-speed of the coupled photon--baryon fluid, and adopting the fitting formula for \(z_{\rm d}\) from Ref.~\cite{Aizpuru:2021vhd}.

\paragraph{Cosmic microwave background.}
We implement the distance-prior compressed likelihood from Ref.~\cite{Zhai:2018vmm}, which encapsulates the essential geometric information of the CMB~\cite{Wang:2007fsa, Zhai:2019nad, Pedrotti:2026ncd}. The compact vector comprises the shift parameter \(R\), the acoustic angular scale \(\ell_{\rm a}\), and the physical baryon density \(\Omega_{\rm b}h^2\), where $h = H_0/(100\, {\rm km}\, {\rm s}^{-1}\, {\rm Mpc}^{-1})$. The sound horizon $r_{\rm s}(z_*)$ required to compute \(\ell_{\rm a}\) is calculated as $r_{\rm d}$ but replacing the redshift at the drag epoch $z_{\rm d}$ by the decoupling redshift $z_*$, which we evaluate using the fitting formula from Ref.~\cite{Aizpuru:2021vhd}.

\paragraph{Type Ia supernovae.}
We consider the full Pantheon\(+\) SN compilation, which contains 1701 light curves spanning the redshift range \(0.001 < z < 2.26\)~\cite{Scolnic:2021amr, Brout:2022vxf}.%
\footnote{The full dataset and covariance matrix are publicly available in the following \texttt{Cobaya} repository: \url{https://github.com/CobayaSampler/sn_data/tree/master/PantheonPlus}.}
We also incorporate the SH0ES calibrated Cepheids~\cite{Riess:2021jrx} to break the well-known \(H_0\)--\(M_{\rm B0}\) degeneracy~\cite{Pan-STARRS1:2017jku}, so that the residual of the apparent magnitude \(\Delta m_i\) is
\begin{equation}
    \Delta m_i =
    \begin{cases}
        m_{{\rm obs},i} - M(z) - \mu_{{\rm Ceph},i}, & i \in {\rm Cepheids},
        \\
        m_{{\rm obs},i} - M(z) - \mu_{{\rm th,i}}, & \rm{otherwise},
    \end{cases}
\end{equation}
with \(\mu_i\) denoting the usual distance modulus, and \(m_{{\rm obs},i}\) the apparent magnitude at the \(i\)th host.

In the absence of screening mechanisms, modifications to the effective gravitational coupling \(G_{\rm eff}\) affect the peak absolute luminosity of supernovae, introducing a redshift dependence on the absolute magnitude \(M(z)\) which can be approximated by~\cite{Riazuelo:2001mg, Gannouji:2018ncm, Kazantzidis:2020tko, Marra:2021fvf}
\begin{equation}
    M(z) =
    M_{\rm B0} - \frac{5}{2} \log_{10} \left(\frac{\mu_G(z)}{\mu_G(0)}\right)^n. \label{eq:sncorr}
\end{equation}
We set \(n = 1.46\) throughout~\cite{Wright:2017rsu, Sakstein:2019qgn, Desmond:2019ygn} instead of the standard Chandrasekhar scaling \(n = -3/2\).

\begin{table}[t]
    \centering
    \caption{Summary of the growth-rate RSD data from various surveys compiled in Ref.~\cite{Euclid:2025bxg}, together with the fiducial (flat \(\Lambda{\rm CDM}\)) matter density values \(\Omega_{\rm m}^{\rm fid}\). We account for the correlations in the WiggleZ block (IDs \(4\)--\(6\)) and the eBOSS block (IDs \(16\)--\(17\)).}
    \label{tab:growth_rate_RSD}
    \begin{ruledtabular}
        \begin{tabular}{clccl}
            ID & \(z_{\rm eff}\) & \(f\sigma_8(z)\) & \(\Omega_{\rm m}^{\rm fid}\) & Reference \\
            \hline
            1 & 0.17 & \(0.510 \pm 0.060\) & \(0.3\) & \cite{Song:2008qt} \\
            2 & 0.02 & \(0.314 \pm 0.048\) & \(0.266\) & \cite{Davis:2010sw} \\
            3 & 0.02 & \(0.398 \pm 0.065\) & \(0.3\) & \cite{10.1111/j.1365-2966.2011.20050.x} \\
            4 & 0.44 & \(0.413 \pm 0.080\) & \(0.27\) & \cite{Blake:2012pj} \\
            5 & 0.60 & \(0.390 \pm 0.063\) & \(0.27\) & \cite{Blake:2012pj} \\
            6 & 0.73 & \(0.437 \pm 0.072\) & \(0.27\) & \cite{Blake:2012pj} \\
            7 & 0.18 & \(0.36 \pm 0.09\) & \(0.27\) & \cite{Blake:2013nif} \\
            8 & 0.38 & \(0.44 \pm 0.06\) & \(0.27\) & \cite{Blake:2013nif} \\
            9 & 1.4 & \(0.482 \pm 0.116\) & \(0.27\) & \cite{Okumura:2015lvp} \\
            10 & 0.02 & \(0.428_{-0.045}^{+0.048}\) & \(0.3\) & \cite{Huterer:2016uyq} \\
            11 & 0.6 & \(0.55 \pm 0.12\) & \(0.3\) & \cite{Pezzotta:2016gbo} \\
            12 & 0.86 & \(0.40 \pm 0.11\) & \(0.3\) & \cite{Pezzotta:2016gbo} \\
            13 & 0.03 & \(0.404_{-0.081}^{+0.082}\) & \(0.312\) & \cite{Qin:2019axr} \\
            14 & 0.013 & \(0.46 \pm 0.06\) & \(0.315\) & \cite{Avila:2021dqv} \\
            15 & 0.15 & \(0.53 \pm 0.16\) & \(0.31\) & \cite{eBOSS:2020yzd} \\
            16 & 0.38 & \(0.500 \pm 0.047\) & \(0.31\) & \cite{eBOSS:2020yzd} \\
            17 & 0.51 & \(0.455 \pm 0.039\) & \(0.31\) & \cite{eBOSS:2020yzd} \\
            18 & 0.70 & \(0.448 \pm 0.043\) & \(0.31\) & \cite{eBOSS:2020yzd} \\
            19 & 0.85 & \(0.315 \pm 0.095\) & \(0.31\) & \cite{eBOSS:2020yzd} \\
            20 & 1.48 & \(0.462 \pm 0.045\) & \(0.31\) & \cite{eBOSS:2020yzd} \\
        \end{tabular}
    \end{ruledtabular}
\end{table}

\paragraph{Growth rate of structure.}
The growth rate of cosmic structure is traced by the \(f \sigma_8\) observable, defined as
\begin{equation}
    f\sigma_8(a) \equiv
    \frac{a\sigma_8}{\delta_{\rm m}(a=1)}\frac{{\rm d}\delta_{\rm m}(a)}{{\rm d}a},
\end{equation}
where \(\sigma_8\) represents the root-mean square fluctuation of the linear matter density field within a comoving sphere of \(8h^{-1}\, {\rm Mpc}\), and \(\delta_{\rm m}\) is the matter density contrast, which obeys
\begin{equation}
    \ddot{\delta}_{\rm m} + 2H\dot{\delta}_{\rm m} =
    4\pi G_{\rm eff} \rho_{\rm m} \delta_{\rm m}.
\end{equation}
The \(f \sigma_8\) can be constrained by using measurements from redshift-space distortion, for which we employ the compilation from Ref.~\cite{Euclid:2025bxg}; see Table~\ref{tab:growth_rate_RSD} for a summary. All growth-rate data are assumed to be uncorrelated, except for the IDs \(4\)--\(6\) block, for which we employ the covariance matrix~\cite{Blake:2012pj}
\begin{equation}
    \boldsymbol{C}_{\rm WiggleZ} = 
    10^{-3} \times
    \begin{pmatrix}
        6.400 & 2.570 & 0.000 \\
        2.570 & 3.969 & 2.540 \\
        0.000 & 2.540 & 5.184  
    \end{pmatrix},
\end{equation}
and for the IDs \(16\)--\(17\) block, for which we employ the covariance matrix~\cite{BOSS:2016wmc}
\begin{equation}
    \boldsymbol{C}_{\rm eBOSS} = 
    10^{-3} \times
    \begin{pmatrix}
        2.034 & 0.812 \\
        0.812 & 1.423
    \end{pmatrix}.
\end{equation}
To minimize the impact of the fiducial (\(\Lambda{\rm CDM}\)) cosmology each survey used to report the \(f\sigma_8(z)\) values in Table~\ref{tab:growth_rate_RSD}, 
we follow Ref.~\cite{Sagredo:2018ahx} and rescale the theoretical prediction for \([f\sigma_8(z)]_{\rm th}\) by the ratios of the \(H(z) D_{\rm A}(z)\) of the cosmology to be constrained to that of the fiducial one, with \(D_{\rm A}(z)\) being the comoving angular-diameter distance.

\paragraph{Solar-System constraints on the gravitational coupling.}
In the absence of screening, the scalar field is light on Solar System scales. Its effective mass is of the order of the Hubble scale, so its range exceeds the astronomical unit by many orders of magnitude~\cite{Esposito-Farese:2000pbo, Argudo-Panes:2026qkw}. The field in the Solar System therefore takes its cosmological background value \(\varphi_0 \equiv \varphi(z=0)\), and local deviations from GR are controlled by \(F(\varphi_0)\) and its derivatives evaluated at \(\varphi_0\). More specifically, these deviations are captured by the PPN parameters~\cite{Boisseau:2000pr, Esposito-Farese:2000pbo}
\begin{equation}
    \gamma - 1 =
    -\frac{F_{\!,\varphi}^2}{F + 2F_{\!,\varphi}^2}, \qquad 
    \beta - 1 =
    \frac{1}{4} \frac{F F_{\!,\varphi}}{2F + 3F_{\!,\varphi}^2} \frac{{\rm d}\gamma}{{\rm d}\varphi}.
\end{equation}
The tightest bound on the present-day value of \(\gamma_0 = \gamma(\varphi_0)\) comes from the Shapiro delay measured with the Cassini spacecraft, \(\gamma_0 - 1 = (2.1 \pm 2.3) \times 10^{-5}\)~\cite{Bertotti:2003rm}, which we include as a Gaussian term in the likelihood. We have not included the additional constraint on \(\beta_0 - 1 = (-2.7 \pm 3.9) \times 10^{-5}\) reported in Ref.~\cite{Park:2017zgd}; nonetheless, we have checked a posteriori that all our reconstruction results are compatible with this test too.

On the other hand, lunar laser ranging constrain the present-day time variation of the gravitational coupling to \((\dot{G}_{\rm eff}/G_{\rm eff}) \vert_0 = (-5.0 \pm 9.6) \times 10^{-15}\, {\rm yr}^{-1}\)~\cite{Biskupek:2020fem}, which we also include as a Gaussian term in the likelihood. Noting that \({\rm d}z/{\rm d}t = -(1 +  z)H\), we compute the latter as
\begin{equation}
    \left.\frac{\dot G_{\rm eff}}{G_{\rm eff}}\right|_0 =
    -1.023\times10^{-10}\,h
    \left.\frac{\mathrm{d}\ln\mu_G}{\mathrm{d}z}\right|_0 \,\mathrm{yr}^{-1}.
\end{equation}

Finally, since \(F(\varphi_{\rm ini}) = 1\) and the field remains frozen during the radiation era, the expansion rate during BBN is set by the bare gravitational coupling entering the action, \(G _{\rm BBN} = G_{\rm N}\), whereas the coupling measured today is \(G_{\rm eff}(0) \equiv G_{\rm 0}\). Their ratio, \(G_{\rm BBN}/G_0 = 1/\mu_G(0)\), is constrained by the primordial light element abundances to \(G_{\rm BBN}/G_0 = 0.99_{-0.05}^{+0.06}\) at \(2\sigma\) confidence level~\cite{Alvey:2019ctk}. We include this as an approximate Gaussian term \(G_{\rm BBN}/G_0 = 0.99 \pm 0.03\) in the likelihood, where, note, we symmetrized the \(1\sigma\) confidence level. Laboratory measurements of the gravitational coupling fix only \(G_{\rm eff}(0)\) and therefore cannot constrain this ratio, which is bounded by BBN alone.

% For titles in APS, download the custom .bst from the following URL:
% https://gist.github.com/goerz/010379874fbff3cc568fbb9fdf63888c
% To activate arXiv links, search for "FUNCTION {format.eprint.controlled}", 
% uncomment the braces below and comment the { "" } before "FUNCTION {format.eprint}"
\bibliographystyle{apsrev4-2-titles}
\bibliography{bibliography}

\end{document}